\documentstyle[12pt]{article}

\input epsf
\input psfig.sty

\begin{document}

\title{Binary recycled pulsars, as a most precise physical
laboratory}

\author{G.S. Bisnovatyi-Kogan
\thanks{Space Research Institute RAN. Profsoyuznaya 84/32, Moscow 117997, Russia; and JINR,
BLTP, Dubna, Russia }}

%\date{}

\maketitle

\begin{abstract}
The following problems are discussed. 1. Pulsars and close binaries.
2. Hulse-Taylor pulsar. 3. Disrupted pulsar pairs. 4.  RP
statistics. 5. Enhanced evaporation: formation of single RP. 6.
General relativity effects: NS+NS. 7. A Double pulsar system. 8.
Checking general relativity. 9. Variability of the gravitational constant.
10. Space Watch.
\end{abstract}
\vspace{1.0cm}

 {\bf PACS numbers:}\quad 97.60.Gb; 95.30.Sf.

 {\bf Key words:}\quad  pulsars; gravitation; precise time.

\section{Introduction}

Discovery of pulsars was announced by  Hewish, Bell et al.\cite{11}.
Until 1973 all known pulsars (more than 100) had been single, while
more than half massive stars (predecessors of pulsars) are in binaries.
Possible explanations of this fact had been suggested: all pairs are
disrupted during explosion, or there is no possibility to form a radiopulsar in pairs
\cite{tr}.

\section{Pulsars and close binaries}

 X-ray satellite UHURU was launched in 1970, and soon
after that X-ray pulsars in
 binaries  had been discovered. One of the best studied
is X ray pulsar Her X-1. It has pulsation period
$ P_p \approx 1.24$ sec,   orbital period  $P_{orb} \approx 1.7$ days,
neutron star mass about $1.4\,M_\odot$, and
optical star with a mass about $2M_\odot$, see e.g \cite{19}.

It was shown in \cite{5}, that this system
should give birth to the binary radiopulsar, by following
reasons.
1. After 100
million years the optical star will become a white dwarf,
 mass transfer will be finished, and the system will be transparent to
 radio emission.
2. X ray pulsar is accelerating its rotation due to accretion, so after
 the birth of the white dwarf, the neurton star will rotate
 rapidly, with $P_p$ about 100 ms.
It was suggested in \cite{5}, that
 binary radiopulsars had not been not found until 1973,
because  the magnetic field of  the neutron star is decreasing about 100 times
during the accretion, so binary radiopulsars are very faint objects.
Pulsar luminosity $L \sim B^2 /P^4$, so
at small B, luminosity L is low even at the rapid rotation.
  Magnetic field is decreasing due to screening by the infalling plasma.

 \section{Hulse-Taylor pulsar}
Hulse and Taylor had discovered \cite{12} the first binary radiopulsar with a period
$P_p=0.05903$ s, orbital period $P_{orb} \approx 7.75$ h, orbit eccentricity $e=0.615$ (see Fig.1).
The properties of the first binary pulsar coincide with our predictions:
rapid rotation together with a small magnetic field \cite{22},\cite{6}
\begin{equation}
\label{ref1} \dot E= -I \Omega\dot\Omega
=I\left(\frac{2\pi}{P_{p}}\right)^2\frac{\dot P_{p}}{P_{p}}
=2\cdot 10^{33}\,{\rm  erg/s}, \quad
\end{equation}
$$ B^2_p=\left(\frac{3Ic^3P_{p}\dot P_{p}}{8\pi^2 R_{NS}^6}\right)^{1/2}
 \approx 2.3\cdot 10^{10}\, {\rm Gs}.$$
The average magnetic fields of single radiopulsars is about $10^{12}$ Gauss.
%%%%%%%%%%%%%%%%%%%%%%%%%%%%%%%%%%%%%%%%%%%%%%%%%%%%%%%%%%%%%%%%
\begin{figure}  %%% FIGURE 2 %%%
\epsfysize=7cm \hspace{1.0cm} \epsfbox{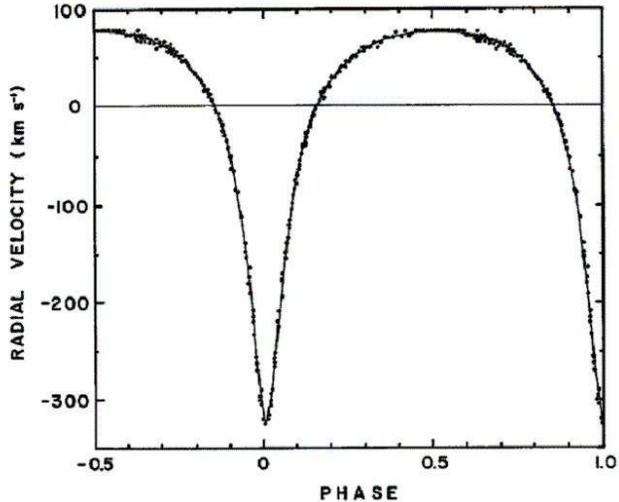} \vspace{0.3cm}
\caption[h]{Velocity curve for the first binary pulsar. Points
represent measurements of the pulsar period distributed over parts
of 10 different orbital periods. The curve corresponds to the pulsar
parameters from the text.}
\end{figure}
%%%%%%%%%%%%%%%%%%%%%%%%%%%%%%%%%%%%%%%%%%%%%%%%%%%%%%%%%%%%%%%%
 \section{Disrupted pulsar pairs}
Suggestions for separation of pulsars at birth was done in \cite{5}
where 10 pairs of single pulsars with a possible origin from one
pair had been listed (see Table 1). The same idea was recently
considered in \cite{25}, who measured pulsar proper motions. They
suggested a common origin of the pair B2020+28 and  B2021+51 (see
Fig.2). It is interesting, that one of this pulsars (2020+28) had
been suggested in \cite{5} for the common origin with another pulsar
(2016+28, line 8 in Table 1). The last pulsar is much closer to
B2020+28 on the sky, and both are situated at the same distance
\cite{lgs98}. The pulsar B2021+51 is much farther on the plane, and
almost two times by the distance, but velocities intersect in Cygnus
OB association.

%%%%%%%%%%%%%%%%%%%%%%%%%%%%%%%%%%%%%%%%%%%%%%%%%%%%%%%%%%%%%%%%
\begin{table}
 \caption{Single pulsars of possible common origin, from \cite{5}.}
\medskip
\begin{tabular}{|l|l|l|l|l|l|l|l|l|l|}
  \hline
No& PSR &R. A.&Decl&$l^{II}$&b$^{II}$&P$_s$&DM $\frac{pc}{cm^3}$&$\dot P\frac{s}{day}$&$\tau=\frac{P}{\dot P}$yr\\ \hline
   1 & P0943+10 & 9$^h$43$^m$20$^s$ &10$^\circ$05$^\prime$33$^{\prime\prime}$ & 225.4 & 43.2 & 1.098 & 15 & - & - \\
     & P0950+08 & 9     50     31     &08        09    43        & 228.9 & 43.7 & 0.253 &  3 & 0.0198 & 3.5$\cdot 10^7$ \\ \hline
   2 & P0809+74 &  8 09  03 & 74  38  10 & 140   & 31.6 & 1.30  & 6 & 0.014 & 2.5$\cdot 10^8$ \\
     & P0904+77 &  9 04     & 77  40     & 135.3 & 33.7 & 1.58  & - & - & - \\ \hline
   3 & P1700-18 & 17 00  56 &-18         &   4.0 & 14.0 & 0.802 &$\le$40 & 0.154 & 6.9$\cdot 10^6$ \\
     & P1706-16 & 17 06  33 &-16  37  21 &   5.8 & 13.7 & 0.653 & 25  & 0.55  & 3.3$\cdot 10^6$ \\ \hline
   4 & P0329+54 &  3 29  11 & 54  24  38 & 145.0 & -1.2 & 0.714 & 27  & 0.177 & 1.1$\cdot 10^7$ \\
     & P0355+54 &  3 55  00 & 54  13     & 148.1 &  0.9 & 0 156 & 55  & - & - \\ \hline
   5 & P0525+21 &  5 25  45 & 21  55  32 & 183.8 & -6.9 & 3.7455& 51  & 3.452  &  3$\cdot 10^6$ \\
     & P0531+21 &  5 31  31 & 21  56  55 & 184.6 & -5.8 &0.03313& 57  & 36.526 & 2.5$\cdot 10^3$ \\ \hline
   6 & P1426-66 & 14 26  34 &-66  09  94 & 312.3 & -6.3 & 0.787 & 60  & - & - \\
     & P1449-65 & 14 49  22 &-65         & 315.3 & -5.3 & 0.180 & 90  & - & - \\ \hline
   7 & P1845-01 & 18 45     &-01  27     &  31.3 &  0.2 & 0.660 & 90  & - & - \\
     & P1845-04 & 18 45  10 &-04  05  32 &  28.9 & -1.0 & 0.598 &142  & - & - \\ \hline
   8 & P2016+28 & 20 16  00 & 28  30  31 &  68.1 & -4.0 & 0.558 & 14.16 & 0.01 & 1.2$\cdot 10^8$ \\
     & P2020+28 & 20 20  33 & 28  44  30 &  68.9 & -4.7 & 0.343 & - & - & - \\ \hline
   9 & P2111+46 & 21 11  41 & 46  36     &  89.1 & -1.2 & 1.015 & 141.4 & - & - \\
     & P2154+40 & 21 54  56 & 40  00     &  90.5 &-11.5 & 1.525 & 110 & - & - \\ \hline
  10 & P0611+22 & 06 11  10 & 22  35     & 188.7 &  2.4 & 0.335 &  99 & - & - \\
     & P0540+23 & 05 40  10 & 23  30     & 184.4 & -3.3 & 0.246 &  72 & - & - \\
  \hline
\end{tabular}
\label{pairpul}
\end{table}

 \section{Recycled pulsar (RP) statistics}

Recycled pulsars is a separate class of neutron stars, containing more than 180 objects.
All passed the stage of accreting pulsars, accelerating the
 rotation and decreasing the magnetic field. So we have
ordinary pulsars with $P_p=0.033$ -- $8$ s, $B= 10^{11}  -  10^{13}$
Gs; and
 recycled pulsars with $P_p=1.5$ -- $50$ ms, $ B=10^8  -  10^{10}$ Gs. These two
 groups are distinctly visible on the $P - \dot{P}$ diagram (Fig.3).

\section{Enhanced evaporation: formation of single RP}

Most recycled pulsars consist of pairs NS + WD, and single NS, about 180
objects in total.
Only 6 NS + NS pairs exist, like Hulse- Taylor
pulsar.
Single recycled pulsars (small $P_p$ and $B$) have passed  the stage of
X-ray pulsar
   in binary, and later had lost the WD companion. Besides,
NS + NS are in the Galaxy disk, and
NS + WD, and single are in the galactic
bulge, and in the globular clusters.
Globular clusters contain 0.001 of the mass of the Galaxy, and about 1/2
 of the recycled pulsars.
Formation of the binary in GC may happen by tidal capture or triple collision.
Disruption of the binary RP also happens by collisions with GC stars: most single RP
 are situated in GC. In the most dense GC we have \cite{15}:
31 RP in Terzian 5 (11 – BP, 10 +10? – single);
22 RP in 47 Tuc  (14 – BP, 7 + 1? – single).
In total:  80 RP in Gal (15 – single),  108 RP in GC (40+15? – single).

Simple disruption of the pair by collisions with field stars does
not work, because hard pairs become even harder by collisions. When
pair is in a state of a disk accretion, with WD filling its Roche
lobe (Fig.4) situation is opposite, and also hard pairs become
softer during collisions due to enhanced mass transfer, and finally
pair is disrupted. This process named "enhanced evaporation", first
considered in \cite{1}, could explain formation of single RP in GC,
and their appearance in bulge may be connected with a full
evaporation of GC (see  also \cite{3}).

\section{General relativity effects:  NS+NS}

\begin{figure}  %%% FIGURE 2 %%%
\epsfysize=8cm \hspace{1.0cm} \epsfbox{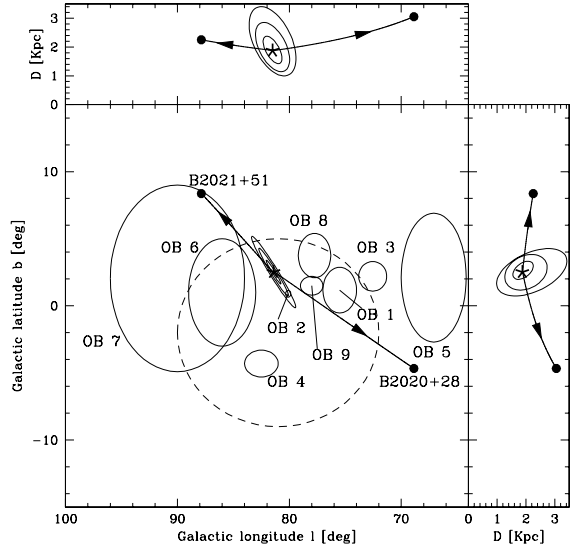} \vspace{0.3cm}
\caption[h]{The 3-dimensional pulsar motion through the Galactic
potential for B2020+28 and B2021+51. The dashed circle represents
the Cygnus superbubble, while the labelled solid ellipses are the
Cygnus OB associations with positions and extents. The extent of OB
2 is unknown and only the centre of the association is indicated.
The thick solid lines indicate the pulsar paths, with the origin
denoted by the starred symbol and the arrows pointing in the
direction of motion. The current positions are indicated by the
solid dots. The elliptical contours around the pulsars' origin in
these panels indicate the 1, 2 and 3$\sigma$ levels of the
likelihood solution for the birth location, from \cite{25}}
\end{figure}

In the NS+NS binary tidal effects are negligibly small, so several post-keplerian (PK)
GR effects had been measured from pulsar timing. They include \cite{t94}
parameter $\dot\omega$,
determining the rate of relativistic apse line motion,
parameter $\gamma$, representing the
amplitude of the signal time delay due to variable gravitational redshift and time
dilation (quadratic Doppler effect), when the pulsar moves in an elliptical orbit.
Emission of gravitational waves results in the loss of the orbital angular
momentum and decreases the orbital period
$\dot{P}_{orb}$. The parameters
$r$ and $s$ determine the time delay due to the
Shapiro effect and are related to the companion's gravitational field.
\begin{equation}
\label{omegadot} \dot{\omega} = 3 T_\odot^{2/3} \; \left(
\frac{P_{orb}}{2\pi} \right)^{-5/3} \;
               \frac{1}{1-e^2} \; (M_A + M_B)^{2/3},
\end{equation}

\begin{equation}
\label{gamma} \gamma  = T_\odot^{2/3}  \; \left( \frac{P_{orb}}{2\pi}
\right)^{1/3} \;
              e\frac{M_B(M_A+2M_B)}{(M_A+M_B)^{4/3}},
\end{equation}

\begin{equation}
\label{pdot} \dot{P}_{orb} = -\frac{192\pi}{5} T_\odot^{5/3} \; \left(
\frac{P_{orb}}{2\pi} \right)^{-5/3} \;
               \frac{\left(1 +\frac{73}{24}e^2 + \frac{37}{96}e^4 \right)}{(1-e^2)^{7/2}} \;
               \frac{M_AM_B}{(M_A + M_B)^{1/3}},
\end{equation}

\begin{equation}
\label{r} r = T_\odot M_B,\,\,\,T_\odot=GM_\odot/c^3=4.925490947\,\, \mu {\rm s},
\end{equation}

\begin{equation}
\label{s} s = x\,T_\odot^{-1/3} \; \left( \frac{P_{orb}}{2\pi}
\right)^{-2/3}
              \frac{(M_A+M_B)^{2/3}}{M_B},
\end{equation}
NS + NS   RPs are the best laboratories for checking of General Relativity.
1913+16 timing had shown (indirectly) the existence of gravitational waves,
Nobel Prize of Hulse and Taylor (1993)
%%%%%%%%%%%%%%%%%%%%%%%%%%%%%%%%%%%%%%%%%%%%%%%%%%%%%%%%%%%%%%%%
\begin{figure}
\centerline { \psfig{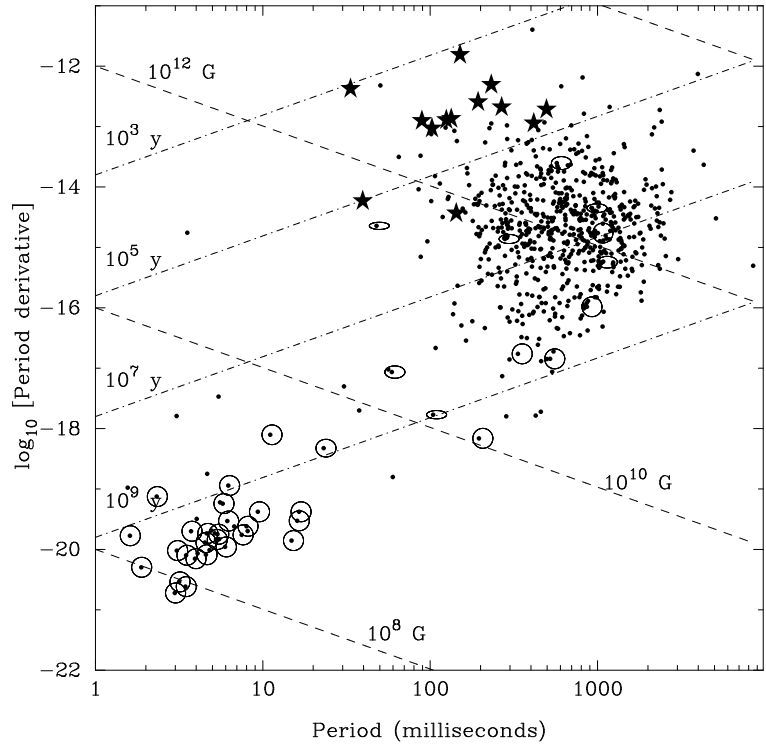} }
\caption{The location of pulsars on the $P -\dot P$ diagram
(period-period derivative). Pulsars in binary systems with
low-eccentricity orbits are encircled, and in high-eccentricity
orbits are marked with ellipses. Stars show pulsars suspected to be
connected with supernova remnants \cite{15}.}
\end{figure}

\begin{figure}[h!]  %%% FIGURE 1 %%%
\epsfysize=8cm \hspace{2.0cm} \epsfbox{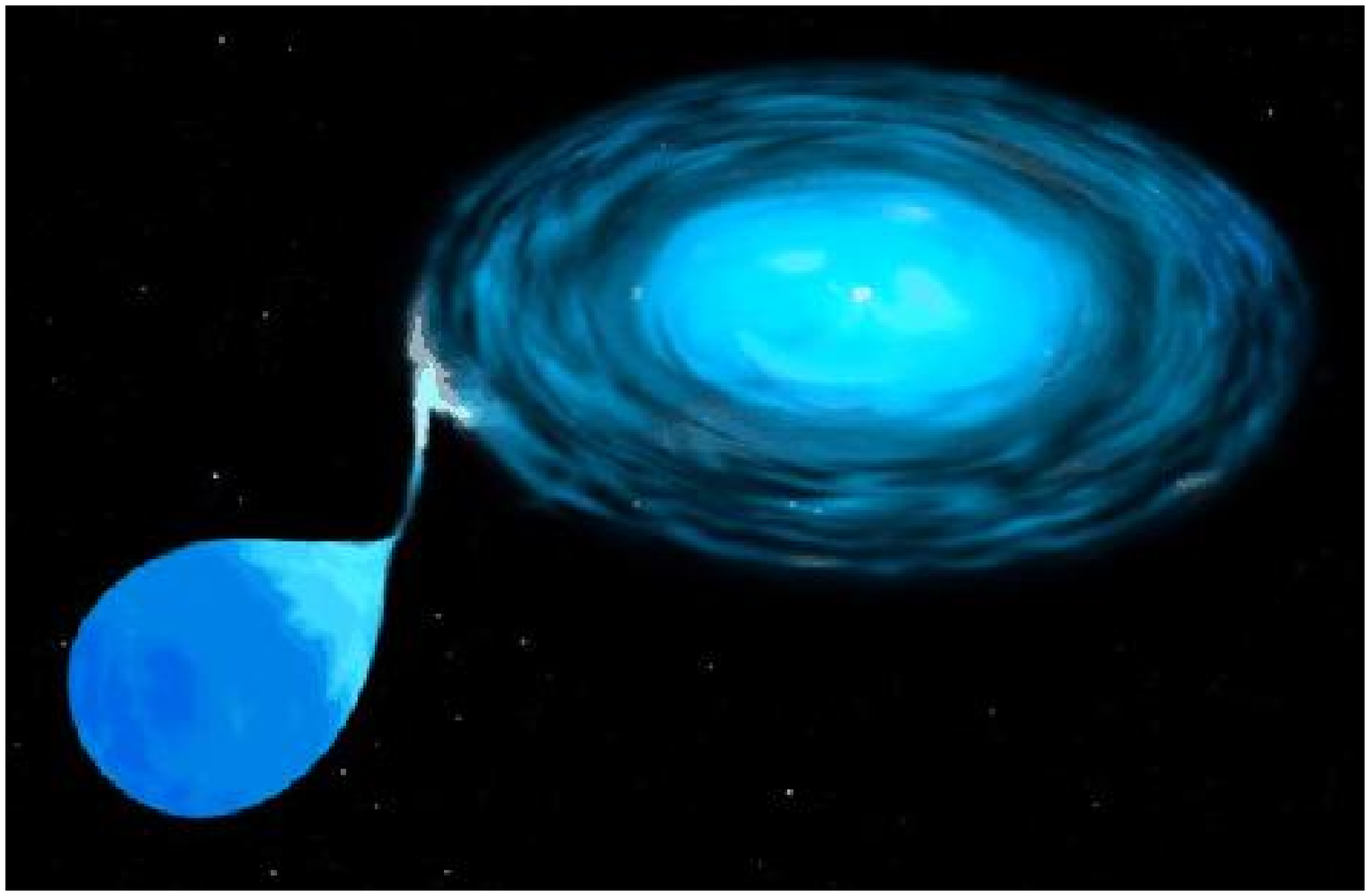} \vspace{0.3cm}
\caption[h]{Artistic vision of a disc accretion to the neutron star
in X-ray sources (http://antwrp.gsfc.nasa.gov/apod/ap991219.html)}
\end{figure}

\section{A double pulsar system}

A binary system, containing two pulsars with periods 23-ms
(J0737-3039A), and 2.8-sec (B) was discovered in 2004 year
\cite{bin04}. This highly-relativistic double-neutron-star system
allows unprecedented tests of fundamental gravitational physics. It
was observed a short eclipse of J0737-3039A by J0737-3039B, and
orbital modulation of the flux density and pulse shape of
J0737-3039B, probably due to the influence of J0737-3039A's energy
flux upon its magnetosphere. These effects will allow to probe
magneto-ionic properties of a pulsar magnetosphere. The
observational properties of this system are listed in Table.2, and
artistic view is give in Fig.5.

\section{Checking general relativity}

 Five equations in the system (\ref{omegadot})-(\ref{s}) contain
two unknown masses. For correct gravitational theory all curves on
the plane $(M_A, \,\, M_B)$ intersect in the same point. Three
curves available for the pulsar 1913+16 (Fig.6), and six curves for
the double pulsar system (Fig.7, the additional line $R$ follows
from observations of both pulsars) intersect in one point inside the
error limits, indicating the correctness of GR on the level
$\sim$0.4\% for 1913+16, after about 20 years of observations, and
on the level $\sim$0.1\% after one year of observations of the
double system.

\section{Variability of the gravitational constant.}

Variations of the gravitational constant $G$ have been measured by different methods
\cite{uzan}. The following restrictions are obtained from
investigations of stellar and planetary orbits.
Using of the Viking Lander ranging data gave limits
$
{\dot G}/{G}=(2\pm 4)\cdot 10^{-12} {\rm yr}^{-1}
$.
The combination of Mariner 10 and Mercury and Venus ranging gave \cite{and92}

\begin{equation}
\label{mariner}
{\dot G}/{G}=(0\pm 2)\cdot 10^{-12} {\rm yr}^{-1},
\end{equation}
Lunar Laser Ranging experiments (LLR) are lasting for many years, and improve the estimations of
${\dot G}/{G}$ from ${\dot G}/{G}< 6\cdot 10^{-12} {\rm yr}^{-1}$
in \cite{dic94}, up to
 $\dot G / G$ [${\rm yr}^{-1}$]= $(6\pm 8)\cdot 10^{-13} $ in
\cite{18}.
Incompletely modeled solid Earth tides, ocean loading or geocenter
motion, and uncertainties in values of fixed model parameters have
to be considered in those estimations.

Recycled pulsars give now the best available timing precision.
The observed changes in the binary period of PSR B1913+16
coincide within error bars with GR prediction of emission of gravitational waves.

%%%%%%%%%%%%%%%%%%%%%%%%%%%%%%%%%%%%%%%%%%%%%%%%%%%%%%%%%%%%%%%%
\begin{figure}[h!]  %%% FIGURE 1 %%%
\epsfysize=6cm \hspace{2.0cm} \epsfbox{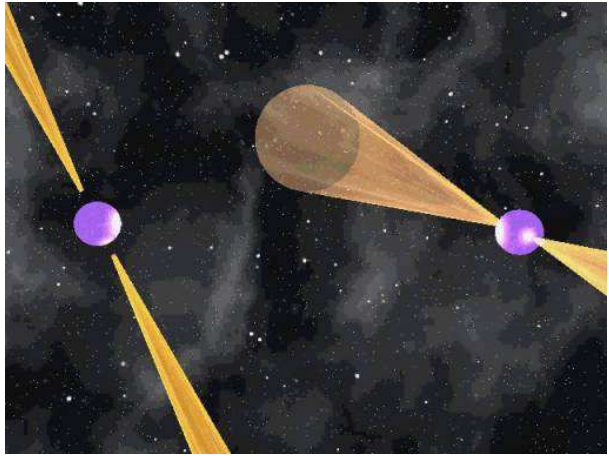} \vspace{0.3cm}
\caption[h]{Artistic picture of the double pulsar system, from
\cite{2}.}
\end{figure}
%%%%%%%%%%%%%%%%%%%%%%%%%%%%%%%%%%%%%%%%%%%%%%%%%%%%%%%%%%%%%%%%
%%%%%%%%%%%%%%%%%%%%%%%%%%%%%%%%%%%%%%%%%%%%%%%%%%%%%%%%%%%%%%%%
\begin{figure}[h!]  %%% FIGURE 1 %%%
\epsfysize=8cm \hspace{2.0cm} \epsfbox{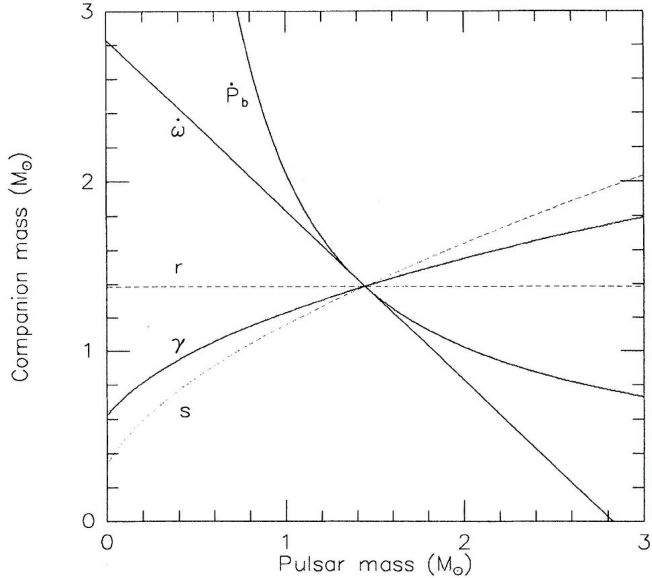} \vspace{0.3cm}
\caption[h]{Observational constraints on the component masses in the
binary pulsar PSR 1913 + 16. The solid curves correspond to Eqns
(2)-(4) with measured values of $\dot\omega$, $\gamma$, and $\dot
P_{orb}$. The intersection of these curves at one point (within an
experimental uncertainty of about 0.35\% in $\dot P_{orb}$) proves
the existence of gravitational waves. The dashed lines correspond to
the predicted values of the parameters r and s. These values can be
measured by future accumulation of observational data \cite{t94}.}
\end{figure}
%%%%%%%%%%%%%%%%%%%%%%%%%%%%%%%%%%%%%%%%%%%%%%%%%%%%%%%%%%%%%%%%
\begin{figure}[h!]  %%% FIGURE 1 %%%
\epsfysize=10cm \hspace{2.0cm} \epsfbox{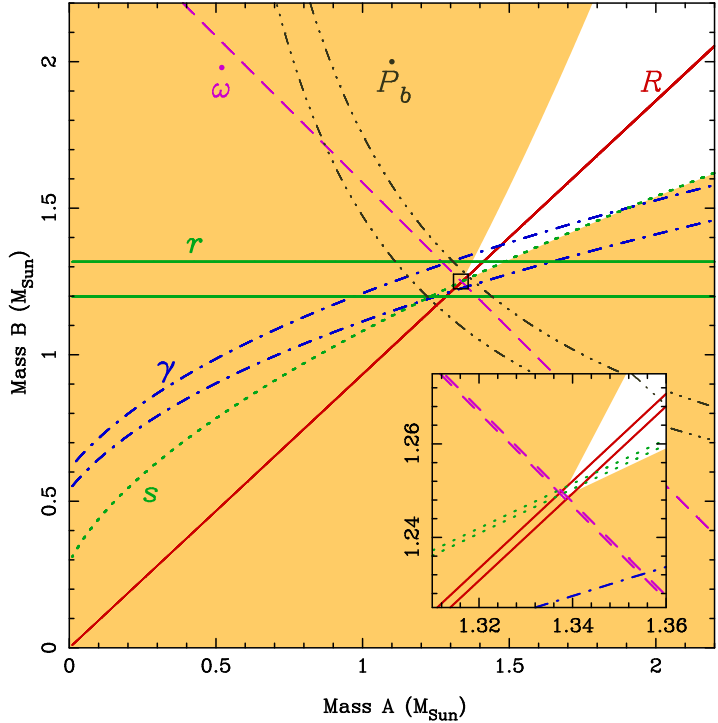} \vspace{0.3cm}
\caption[h]{The observational constraints upon the masses mA and mB
in the double pulsar system J0737--3039.. The colored regions are
those which are excluded by the Keplerian mass functions of the two
pulsars. Further constraints are shown as pairs of lines enclosing
permitted regions as predicted by general relativity: (a) the
measurement of the advance of periastron, giving the total mass
mA+mB = 2:588 pm 0:001M (Sun) (dashed line); (b) the measurement of
R = mA/mB = 1:071 $\pm$ 0:001 (solid line); (c) the measurement of
the gravitational redshift/time dilation parameter (dot-dash line);
(d) the measurement of Shapiro parameter r (dot-dot-dot-dash line)
and (e) Shapiro parameter s (dotted line). Inset is an enlarged view
of the small square encompassing the intersection of these
constraints, from \cite{bipgr05}.}
\end{figure}
\noindent
Measurements of the variations
of the binary period give restrictions
to the variation of the gravitational constant. Change of $G$ has influence on the
orbital motion, and only emission of the gravitational waves compete here with the variation of
$G$.
If we accept the correctness of GR,
than we can use residuals (error box) of the measurements of
decay of the binary period $\dot P_b$ for the estimation of variations of $G$
\cite{dam88}. It was obtained in \cite{kaspi},
using timing data from \cite{20}
\begin{equation}
\label{kasp}
\frac{\dot G}{G}=-\frac{1}{2}\frac{\dot P_b}{P_b}
=(4\pm 5)\cdot 10^{-12} {\rm yr}^{-1}.
\end{equation}
For the pulsar PSR B1913+16 the error budget for the orbital period
derivative, in comparison with GR prediction, is given in Table 3
\cite{20}. Supposing that all deviations from GR are connected with
$G$ variation, we obtain the upper limit for these variation as
follows \cite{4}: $ \frac{\dot G}{G}=(4.3\pm 4.9)\cdot 10^{-12}\,
{\rm yr}^{-1}. $
\begin{table}
\caption{Observed and derived parameters of PSRs~J0737$-$3039A and B.
Standard errors are given in parentheses after the values and are in
units of the least significant digit(s).
The distance is estimated from the dispersion measure and a
model for the interstellar free electron distribution,
from \cite{bin04},\cite{bipgr05}.}
\footnotesize
\begin{tabular}{lcc}
\hline
\hline
 & & \\
Pulsar & PSR~J0737$-$3039A & PSR~J0737$-$3039B \\
Pulse period $P$ (ms) & 22.699378556138(2) & 2773.4607474(4) \\
Period derivative $\dot{P}$ & $1.7596(2) \times 10^{-18}$ & $0.88(13)\times 10^{-15}$ \\
Right ascension $\alpha$ (J2000) &
  \multicolumn{2}{c}{$07^{\rm{h}}37^{\rm{m}}51^{\rm{s}}.24795(2)$ } \\
Declination $\delta$ (J2000) &
  \multicolumn{2}{c}{$-30^\circ 39' 40''.7247(6)$ } \\
Orbital period $P_{\rm{b}}$ (day) &
  \multicolumn{2}{c}{0.1022515628(2)} \\
Eccentricity $e$ & \multicolumn{2}{c}{0.087778(2) } \\
Advance of periastron $\dot{\omega}$ (deg yr$^{-1}$) &
  \multicolumn{2}{c}{16.900(2)}\\
Projected semi-major axis $x=a{\rm{sin}}i/c$ (sec) & 1.415032(2) & 1.513(4) \\
Gravitational redshift parameter $\gamma$ (ms) & 0.39(2) &  \\
Shapiro delay parameter $s=\sin i$ & $0.9995(4)$ & \\
Shapiro delay parameter $r$ ($\mu$s) & $6.2(6)$ &  \\
Orbital decay $\dot{P}_b$ ($10^{-12}$) & $-1.20(8)$ \\
Mass ratio $R=M_A/M_B$ & \multicolumn{2}{c}{1.071(1) } \\
Characteristic age $\tau$ (My) & 210 & 50 \\
Surface magnetic field strength $B$ (Gauss) & $6.3\times 10^9$ & $1.6\times 10^{12}$ \\
Spin-down luminosity $\dot E$ (erg/s) & $5800\times10^{30}$ & $1.6\times10^{30}$ \\
Distance (kpc) & \multicolumn{2}{c}{$\sim$0.6} \\
%Geodetic precession rate (deg yr$^{-1}$) & 4.79 & 5.07 \\
 & & \\
\hline
\end{tabular}
\end{table}
These estimations
slightly improve the values \cite{kaspi}.
This improvement narrows on 40\% the boundary in the negative region, most important from the
theoretical point of view.
Combination the last result with \cite{and92} permits
to narrow the range of $G$ variations, which is now may be situated in the limits
\cite{4}

$${\dot G}/{G}\,\, {\rm is\,\, within \,\,the \,\,interval}
\,\,(-0.6, \,\,2)\cdot 10^{-12} {\rm yr}^{-1}.
$$

\begin{table}
 \caption{Error budget for the orbital period derivative,
in comparison with the general relativistic prediction.}
\medskip
\begin{tabular}{lll}
%\tableline
                       & Parameter  &($10^{-12}$)   \\
                       \hline
 Observed value        &  $\dot P_b^{obs}$  & -2.4225$\pm$0.0056   \\
 Galactic contribution & $\dot P_b^{gal}$   & -0.0124$\pm$0.0064  \\
 Intrinsic orbital period decay & $\dot P_b^{obs}$ - $\dot P_b^{gal}$&-2.4101$\pm$0.0085  \\
 General relativistic prediction & $\dot P_b^{GR}$ &  -2.4025$\pm$0.0001     \\
\hline \hline
\end{tabular}
\label{htpul}
\end{table}

\section{MATRE: Most Accurate Time measurements by REcycled pulsars
(Space Watch)}

Exact time measurements are needed in many branches of human activity,
as well as by simple pedestrians. The nature had prepared us the most
accurate available watch in the form of Recycled Pulsars. The precision
of the period of these pulsars is very high, and they are changing their
periods with time incredibly slowly: $dP/dt=10^{-19.5}$ for
PSR 1953+29, P=6.133 ms; $dP/dt=10^{-19.0}$ for PSR 1937+21, P=1.558 ms,
what is, for a long time period, much better  than the highest available
precision based on the hydrogen frequency standard.

    I suggest to build the international time watch system based on radio
    observations of several most stable recycled pulsars. To build this
    system the following components should be constructed (see fig.9).

\begin{figure}
\centerline { \psfig{figure=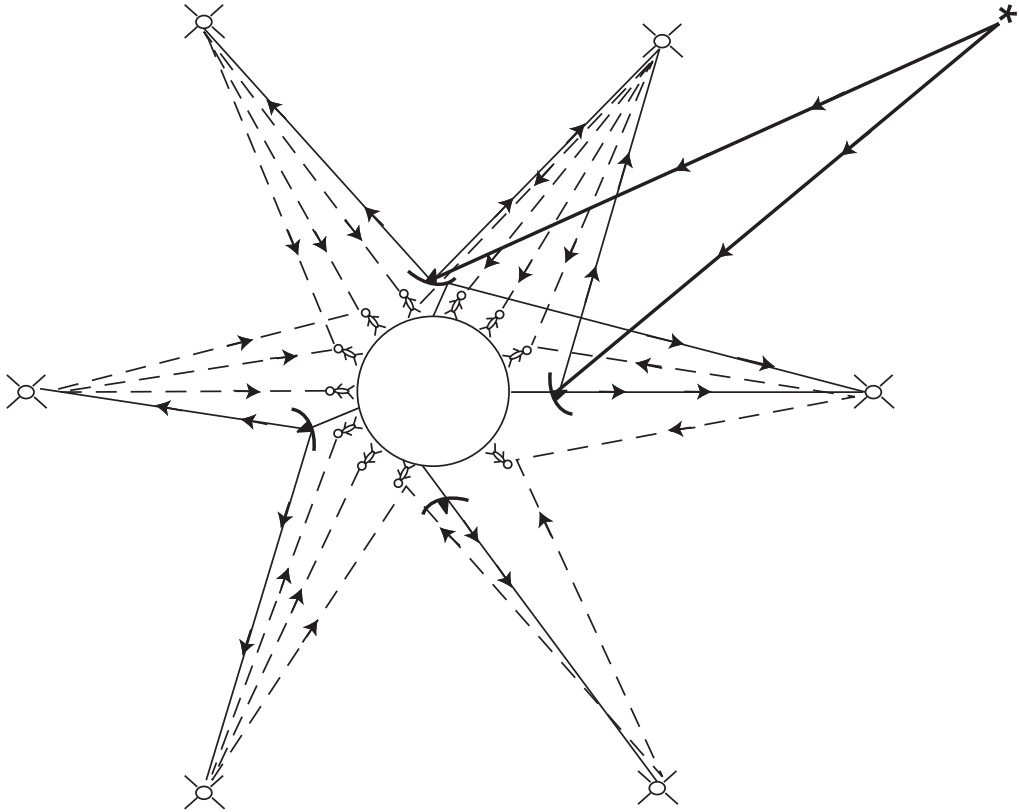,width=14cm,angle=-90} }
\caption{Schematic sketch of the Space Watch system. The signal from
the recycled pulsar (upper left) is accepted by the Earth radio
telescopes, which transfer it to the satellite system. The
pedestrians accept the exact time signal from the satellite by
receivers similar to those used in GPS (or GLONASS).}
\end{figure}

    1. Several radio telescopes over the Earth, which provide stable
    continuous observations of the chosen pulsar(s) during 24 hours,
    and retransmission of the signal to the Earth satellites.

    2. System of the Earth satellites (20-25) which accept the
    exact time signals from the radio telescopes, and retransmit them
    to the Earth, covering all the Earth surface, similar to GPS system.

    3 Device which accept the exact tine signal from the satellite,
    and give the exact local/universal time information (should be
    widely available, could be combined with the cell phone, or GPS device).).

    4. Software/hardware for transformation of the pulsar signal into
    the signal of the exact time measurement on the radio telescopes,
    satellites, and in every receiving device.

    Such system could be used in all branches of the human activity,
    including ordinary citizen. It should be really international,
    because its work depends on the synchronous and non-interruptive
    work of all telescopes in different countries. The cost of the
    construction of this system should be quite moderate (comparable
    with the construction of GPS), but the exact time is needed in much
    more wide cases than exact coordinates, simply because overwhelming
    majority of people have permanent place of life, or are moving along
    the well known routes. But watches are used by all of them. Therefore,
    this system could be quite profitable with time, because after
    construction it  needs not much for supporting its work.

\section{Conclusions}

1. Timing of the pulsars J0737-3039A /B is the most powerful
 instrument for the verification of General Relativity
   due to unprecedented precision of the observations.
2. Recycled pulsars are the most precise available time standards.
3. Checking the physics beyond  the standard: G-variability is possible by
   timing of recycled pulsars in close binaries.
4. The service of the exact time based on continuous observations of
   recycled pulsars may be constructed and used by pedestrians.

\medskip

{\bf Acknowledgements}
This work was partially supported by RFBR grants 05-02-17697A, 06-02-91157 and
06-02-90864, and President grant
for a support of leading scientific schools 10181.2006.2..

\end{document}